\documentclass[aps,prl,twocolumn,superscriptaddress,floatfix]{revtex4-1}

\usepackage{amsmath,amssymb}
\usepackage{graphicx}
\usepackage{color}
\usepackage{hyperref}

\begin{document}


\title{All-electric single electron spin initialization}


\author{S. Bednarek}
\affiliation{
	Faculty of Physics and Applied Computer Science,
	AGH University of Science and Technology, Krak\'{o}w, Poland}
\author{J. Paw\l{}owski}
\email[]{jaroslaw.pawlowski@pwr.edu.pl}
\affiliation{
Department of Theoretical  Physics, Faculty of Fundamental Problems of Technology, Wroc\l{}aw University of Science and Technology, Wybrze\.{z}e Wyspia\'{n}skiego 27, 50-370 Wroc\l{}aw, Poland}
\author{M. G\'orski}
\affiliation{
	Faculty of Physics and Applied Computer Science,
	AGH University of Science and Technology, Krak\'{o}w, Poland}	
\author{G. Skowron}
\affiliation{
	Faculty of Physics and Applied Computer Science,
	AGH University of Science and Technology, Krak\'{o}w, Poland}
	
\date{\today}

\begin{abstract}
We propose a nanodevice for single-electron spin initialization. It is based on a gated planar semiconductor heterostructure with a quantum well and with potentials generated by voltages applied to local gates. Initially we insert an electron with arbitrary spin into the nanodevice. Next we perform a sequence of spin manipulations, after which the spin is set in a desired direction (e.g., the growth direction). The operations are done all-electrically, do not require any external fields and do not depend on the initial spin direction.
\end{abstract}

\pacs{}

\maketitle

Since the emergence of first quantum algorithms, which have shown that there exist problems, that can be solved much faster by a quantum computer \cite{alg1,alg2}, an intense research on quantum computing and its possible physical implementations have begun. Qubits can be defined by nuclear spins of molecules, operated on using NMR \cite{nmr3,nmr4,nmr5}, internal states of isolated atomic ions \cite{trap6,trap7,trap8,trap9} or spins of electrons (holes) trapped in quantum dots (QD), or more generally, quantum semiconductor nanostructures \cite{spin10,spin11,spin12}.

Regardless of the choice of a qubit carrier, one has to be able to perform the most fundamental operations \cite{uniwers13,uniwers13b}: initialization, manipulation and readout. For each qubit implementation, these operations require appropriate methods, which affect the fidelities and duration of operations. Fidelity near $100\%$ is being achieved for qubits realised with ion traps \cite{trap7,trap8}, for which the initialization and manipulation are performed using pulses of light. In QDs, in which confinement is created by heterojunctions of semiconductors, spin can also be controlled with photons. In self-assembled QDs, it is possible to initialize spin of an electron \cite{hetero14,hetero15,hetero16} or a hole \cite{hetero18,hetero19,hetero20,hetero21,hetero22} along the direction of magnetic field using optical transitions through trionic states. Similarly spin can be initialized in QDs formed by heterojunctions in catalytically grown nanowires \cite{hetero17}. The achieved fidelities exceed 99\% \cite{hetero15,hetero17,hetero19,hetero20}.

In electrostatic QDs the lateral confinement is generated in a quantum well (QW) \cite{elst23,elst24,elst25,elst26,elst27} or a quantum wire \cite{elst28,elst29} using voltages applied to gates. In such systems it is impossible to initialize spin optically through trionic states, since the electrostatic potential, which is attractive for electrons, is repulsive for holes and cannot create a stable excitonic state. Instead, spin is initialized using Pauli blockades \cite{elst26,blokada31}. This method allows to set spin of the second electron in a double QD in the same direction as spin of the previously trapped electron. To deliberately set spin of the first electron, one has to apply a strong magnetic field and wait until the electron relaxes to the ground state \cite{elst27}. Unfortunately, this operation is neither exact nor fast. In the literature we can find theoretical proposals of fast spin initialization \cite{fast32,fast33,fast34} but none has been verified experimentally. In our article, we present a nanodevice, which allows to initialize the electron spin in a few hundred picoseconds with fidelity exceeding 99\%.

In this work we propose a nanodevice capable of initializing spin of a single electron using exclusively the electric field. As a quantum bit carrier we assume spin of a single electron confined inside a QW. The qubit basis states correspond to spin states with orientation parallel and antiparallel to the z-axis. During operation of the proposed nanodevice, an arbitrary spin state of the electron is turned into a state with spin parallel to the $z$-axis. The entire process is divided into two stages. In the first stage we separate the electron wavepacket into two parts of opposite spins (parallel to the $y$-axis in the left half of the nanodevice and antiparallel in the right). In the second stage, due to spatial separation of both parts, we rotate their spins independently in opposite directions by $90^\circ$. As a result, spins of both parts become parallel to the $z$-axis, which is the goal of the spin initialization procedure.

The proposed device is based on a planar semiconductor heterostructure with a QW parallel to its surface. We assume InSb for the QW material due to its strong spin-orbit coupling. The potential barriers for the QW are created by presence of two adjacent layers of $\mathrm{Al_xIn_{1-x}Sb}$ on both sides of the QW. It is so, because for the assumed $x=25\,\%$ the bottom of the conduction bands in $\mathrm{Al_xIn_{1-x}Sb}$ and InSb are shifted by about $300\,\mathrm{meV}$ \cite{offset1,offset2,offset3} with respect to one another. The growth direction of the heterostructure must be chosen as the $[111]$ crystallographic direction. In such a case the Rashba and Dresselhaus spin-orbit interactions (SOI) are described by operators of the same form and can be merged \cite{(111)}.
\begin{figure}[h]
\centering
\includegraphics[width=0.45\textwidth,trim={0 0cm 0 0.5cm}]{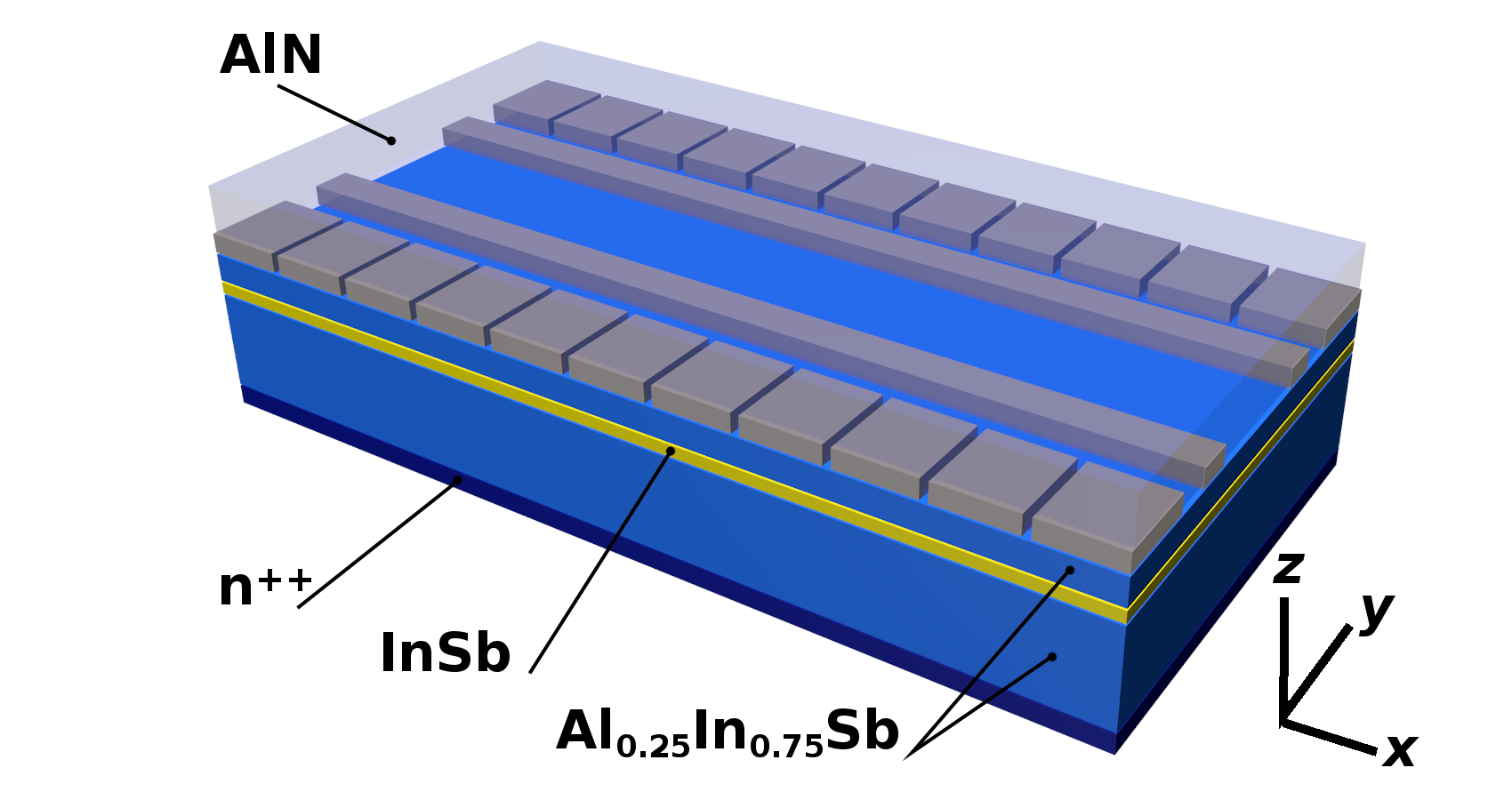}
\caption{\label{fig:structure}Schematic view of the considered nanodevice.}
\end{figure}

\begin{figure}[h]
\centering
\includegraphics[width=0.45\textwidth]{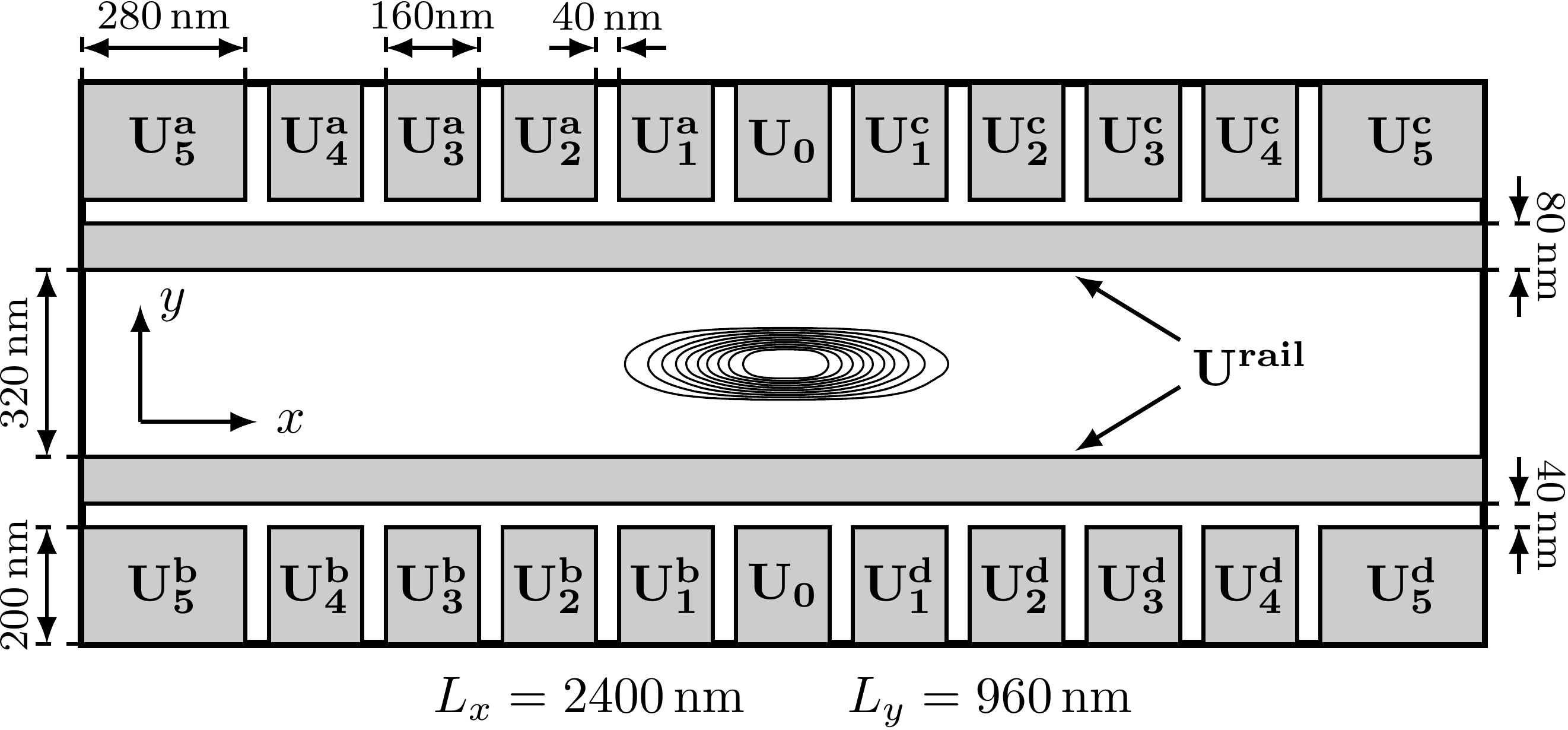}
\caption{\label{fig:layout}Layout of gates and the initial charge density.}
\end{figure}

On the nanostructure substrate, which consists of highly donor-doped $\mathrm{Al_xIn_{1-x}Sb}$ ($\mathrm{n^{++}}$), we first deposit a $230\,\mathrm{nm}$ wide barrier layer of $\mathrm{Al_xIn_{1-x}Sb}$, next a $20\,\mathrm{nm}$ wide InSb layer constituting the QW and a second $50\,\mathrm{nm}$ wide barrier layer of $\mathrm{Al_xIn_{1-x}Sb}$ (see Fig. \ref{fig:structure}).  On this layer we place metallic gates (see Fig. \ref{fig:layout}). The entire nanodevice is covered with an additional $100\,\mathrm{nm}$ wide dielectric layer of $\mathrm{AlN}$. Finally we cover the top of the dielectric with another metallic gate, subsequently referred to as top gate $U^\mathrm{top}$ (not shown in the figures).

If no external electric fields are present, electrons from the substrate fill the QW forming a two-dimensional electron gas (2DEG). Electrons trapped in the QW have two motional degrees of freedom $(x,y)$. If we now apply appropriate negative voltages to the gates, the gas can be depleted until only one electron remains. It is confined between two inner gates $U^\mathrm{rail}$ subsequently referred to as rails. The rails can block electron movement along the $y$ direction and define a path along which the electron can move. Note that all voltages are applied with respect to the substrate.

Initially, we apply a voltage $U_{\mathrm{bias}}=-400\,\mathrm{mV}$ to the top gate $U^\mathrm{top}$ and the rails $U^\mathrm{rail}$. The same voltage is applied to two lateral gates denoted by $U_0$, hence we have $U_0(t=0)=U_{\mathrm{bias}}$. Voltages applied to remaining gates grow proportionally to the square of the distance from the center of the nanodevice calculated along the $x$-axis. Since the gates are of equal widths, this translates directly into proportionality to the square of the gate index. Thus we have $U_i^{\mathrm{a,b,c,d}}(t=0)=U_{\mathrm{bias}}+i^2U_{\mathrm{par}}$ with $U_{\mathrm{par}} = -5\,\mathrm{mV}$. The potential inside the nanodevice is calculated using the generalized Poisson's equation \cite{pawlowski1} at every time step \cite{metoda}. The obtained potential takes into account applied variable voltages, the device geometry and time-dependent charge distribution inside the QW \cite{metoda,poiss1}. It also includes the image charge induced on the gates, self-focusing the electron wavefunction, as well as dielectric polarization counteracting this effect \cite{naszkot}. In this way we can generate a nearly perfect parabolical confinement potential along the $x$-axis, which is necessary to generate coherent states of the harmonic potential \cite{trap9}.

The electron confined in the QW has two motional degrees of freedom $(x,y)$ and its Hamiltonian takes the form
\begin{equation}\label{eq:hamiltonian}
\mathbf{\hat{H}}=\left[\frac{-\hbar^2}{2m}\nabla^2-|e|\varphi_{\mathrm{el}}(x,y,z_0,t)\right]\mathbf{I}_2+\mathbf{\hat{H}}_{\mathrm{SO}},
\end{equation}
where $\nabla^2=\partial_{x}^2+\partial_{y}^2$ and $m=0.014m_e$ is the electron effective mass in InSb, $-|e|\varphi_{\mathrm{el}}$ is the confinement potential felt by the electron and $z_0$ is the $z$-coordinate of the QW plane. The potential is calculated at every time step using the Poisson's equation with boundary conditions taking into account voltages applied to the gates \cite{metoda}. Note that $\mathbf{I}_2$ in Eq. (\ref{eq:hamiltonian}) is a $2\times 2$ identity matrix. The wavefunction takes the form of a spinor $\mathbf{\Psi}= \big(\Psi_\uparrow(x,y,t),\Psi_\downarrow(x,y,t)\big)^T$.

The last term in (\ref{eq:hamiltonian}) is the sum of the Rashba and Dresselhaus SOI of the following form
\begin{equation}\label{eq:soihamiltonian}
\mathbf{\hat{H}}_{\mathrm{SO}} = (\alpha(x,y,t)+\beta)(\sigma_x\hat{p}_y-\sigma_y\hat{p}_x),
\end{equation}
with Pauli matrices $\sigma_x, \sigma_y$. The Dresselhaus coupling $\beta=\gamma(\pi/d)^2/\hbar$ is calculated for the QW width $d=20\,\mathrm{nm}$ and Dresselhaus coefficient for InSb $\gamma=228.3\,\mathrm{eV\,\AA{}^3}$ \cite{DresInSb1,DresInSb2,DresInSb3}. The Rashba coupling is calculated for the $z$-component of the electric field $E_z$ within the QW as $\alpha(x,y,t)=\alpha_{\mathrm{SO}}|e|E_z(x,y,z_0,t)/\hbar$, with Rashba coefficient $\alpha_{\mathrm{SO}}=523\,\mathrm{\AA{}^2}$ adequate for InSb \cite{winkler}. 

For the initial state of the electron trapped in our nanodevice we assume its ground state in the confinement potential. We examine the time evolution of the electron wavefunction solving the time-dependent Schr\"odinger equation iteratively. To include time-dependency of voltages and the charge distribution in the QW we solve the Poisson's equation at every time step \cite{metoda,pawlowski1}. During the first phase of time evolution we change all the gate voltages sinusoidally according to the formula $U_i^{\mathrm{a,b,c,d,rail,top}}(t)=U_i^{\mathrm{a,b,c,d,rail,top}}(0)+\Delta U\sin(\Omega t)$, with the amplitude $\Delta U=350\,\mathrm{mV}$ and frequency of oscillations $\Omega$ tuned to the natural frequency of the harmonic confinement potential along the $x$-axis. In our simulations the optimal value is $70\,\mathrm{GHz}$ ($\hbar\Omega=0.2895\,\mathrm{meV}$). At this stage of simulation we shift all the voltages by the same value, thus the shape of the potential along the $x$-axis does not change over time but the level of its bottom does. 

Oscillating voltages applied to the gates result in oscillations of $E_z$ in the QW, which in turn introduces oscillations to the Rashba coupling\cite{nitta}. According to the SOI Hamiltonian (\ref{eq:soihamiltonian}), such oscillations can induce electron motion in the $x$ and $y$ directions. However, negative voltages applied to the rails ensure strong confinement in the $y$-direction, hence the amplitude of oscillations in this direction is very small. Motion along the $x$-axis is less restrained. If the electron spin is parallel or antiparallel to the $y$-axis, the electron starts to oscillate in the coherent state of the harmonic oscillator \cite{trap9} with increasing amplitude \cite{naszkot}. If spin is parallel (antiparallel) to the $y$-axis, the change of the SOI coupling initially push the electron to the right (left). As a result electrons with $\langle\sigma_y\rangle=1$ and $\langle\sigma_y\rangle=-1$ oscillate with opposite phases.
\begin{figure}[h]
\centering
\includegraphics[width=0.45\textwidth]{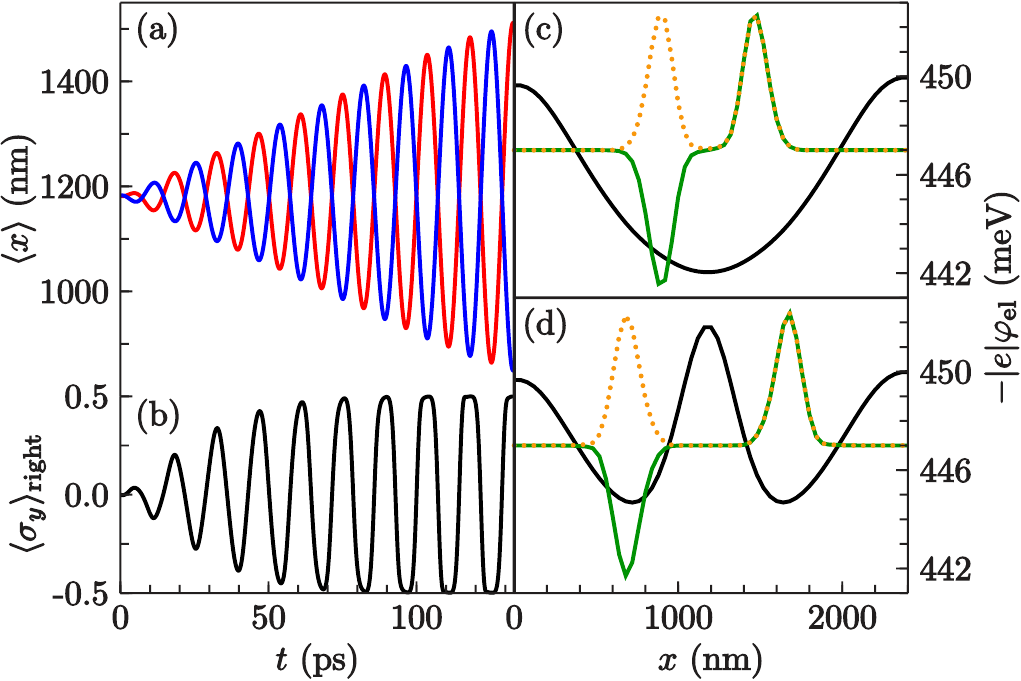}
\caption{\label{fig:rashbaosc} $a)$ Expectation values of the electron position along the $x$-axis for $\langle\sigma_y\rangle=1$ (red) and $\langle\sigma_y\rangle=-1$ (blue). $b)$ Expectation values of spin $\langle\sigma_y\rangle_{\mathrm{right}}$ calculated for the right half of the nanodevice (for $x>L_x/2$) according to the expression (\ref{eq:spinyright}). $c,d)$ Electron (dotted orange) and spin (solid green) densities just before and after setting up an additional barrier between two separated spin parts. The black curve shows a potential profile along the $x$-axis.}
\end{figure}
The red and blue curves (Fig. \ref{fig:rashbaosc}a) show the $x$-component of the expectation value of position of the electron $\langle x\rangle$ for spins parallel and antiparallel to the $y$-axis, respectively.

Now, let us assume that the spin is set in parallel to the $z$-axis. The spin wavefunction $\chi_z$ can be expressed as a linear combination of wavefunctions corresponding to spin parallel and antiparallel to the $y$-axis denoted as $\chi_y$ and $\chi_{-y}$ respectively 
\begin{equation}\label{eq:spinz}
\chi_z=\begin{pmatrix}1\\0\end{pmatrix}=\frac{1}{2}\begin{pmatrix}1\\i\end{pmatrix}+\frac{1}{2}\begin{pmatrix}1\\-i\end{pmatrix}=\frac{1}{\sqrt{2}}(\chi_y+\chi_{-y}).
\end{equation}
For this state $\langle\sigma_y\rangle=0$. Generally we can express the full wavefunction as
\begin{equation}\label{eq:fullspinz}
\mathbf{\Psi}(x,y,t)=\Psi_y(x,y,t)\chi_y+\Psi_{-y}(x,y,t)\chi_{-y}.
\end{equation}
Let us now define the electron density $\rho=\rho(x,t)$ and spin density $\rho_\sigma=\rho_\sigma(x,t)$ (along the $x$-axis) as
\begin{equation}\label{eq:chargedensity}
\rho=\int_0^{L_y}dy\mathbf{\Psi}^\dagger\mathbf{\Psi}=\int_0^{L_y}dy\left(|\Psi_y|^2+|\Psi_{-y}|^2\right),
\end{equation}
\begin{equation}\label{eq:spindensity}
\rho_\sigma=\int_0^{L_y}dy\mathbf{\Psi}^\dagger\sigma_y\mathbf{\Psi}=\int_0^{L_y}dy\left(|\Psi_y|^2-|\Psi_{-y}|^2\right).
\end{equation}
Given these definitions we can describe the electron behavior.

Initially both spatial components $\Psi_y$ and $\Psi_{-y}$ are identical and equal to the spatial wavefunction of the initial state of the electron---being the ground state in the confining potential. During time evolution they behave accordingly to their spin and their $\langle x\rangle$ oscillate as shown in Fig. \ref{fig:rashbaosc}a. The black curve in Fig. \ref{fig:rashbaosc}b shows $\langle\sigma_y\rangle$ calculated only for the right half of the nanodevice according to the formula
\begin{equation}\label{eq:spinyright}
\langle\sigma_y\rangle_{\mathrm{right}}=\int_{L_x/2}^{L_x}dx\rho_\sigma(x,t).
\end{equation}

Initially $\langle\sigma_y\rangle_{\mathrm{right}} = 0$, because both spatial components of the wavefunction are identical. After a while they no longer overlap and if $\Psi_y$ is shifted to the right ($\Psi_{-y}$ to the left), the value of $\langle\sigma_y\rangle_{\mathrm{right}}$ is positive. For an opposite shift, $\langle\sigma_y\rangle_\mathrm{right}$ is negative. Note that the amplitude of oscillations of $\langle\sigma_y\rangle_{\mathrm{right}}$ grows until it reaches $0.5$. The value of $0.5$ indicates full spatial separation of both spin components of the wavefunction. Fig. \ref{fig:rashbaosc}c shows the wavepacket at this very moment. The dotted orange and solid green curves denotes the electron and spin densities along the $x$-axis, $\rho$ and $\rho_\sigma$ respectively. According to (\ref{eq:chargedensity},\ref{eq:spindensity}), in the area where the spin is parallel to the $y$-axis both densities overlap, while in the area with antiparallel spin the densities have opposite signs. Fig. \ref{fig:rashbaosc}c shows a situation with spin parallel to the $y$-axis in the right half of the nanodevice and antiparallel in the left. This is a Schr\"odinger's cat-like state. A similar state has been obtained experimentally in ion traps \cite{trap9}. 

Now if we cease the oscillations of voltages and rise a potential barrier between two parts of the wavefunction of opposite spins, we can separate them permanently. We achieve this by lowering voltages applied to $U_0$ by $875\,\mathrm{mV}$. The obtained potential and the electron and spin densities are visible in Fig. \ref{fig:rashbaosc}d. The potential energy has two minima separated by a barrier. It has to be high enough and the minima sufficiently deep to allow for operations on both spin parts independently. The voltages are switched fast but the switching moment should be chosen, so that the positions of the potential minima coincide with centers of both wavepacket parts. This way we avoid an unnecessary rise of the electron energy. We thus obtained a state with spatially separated spin parts. If we start the separation from a different superposition of states parallel and antiparallel to the $y$-axis, full separation will take the same amount of time but the final value of $\langle\sigma_y\rangle_{\mathrm{right}}$ will be different.

\begin{figure}[h]
\centering
\includegraphics[width=0.45\textwidth]{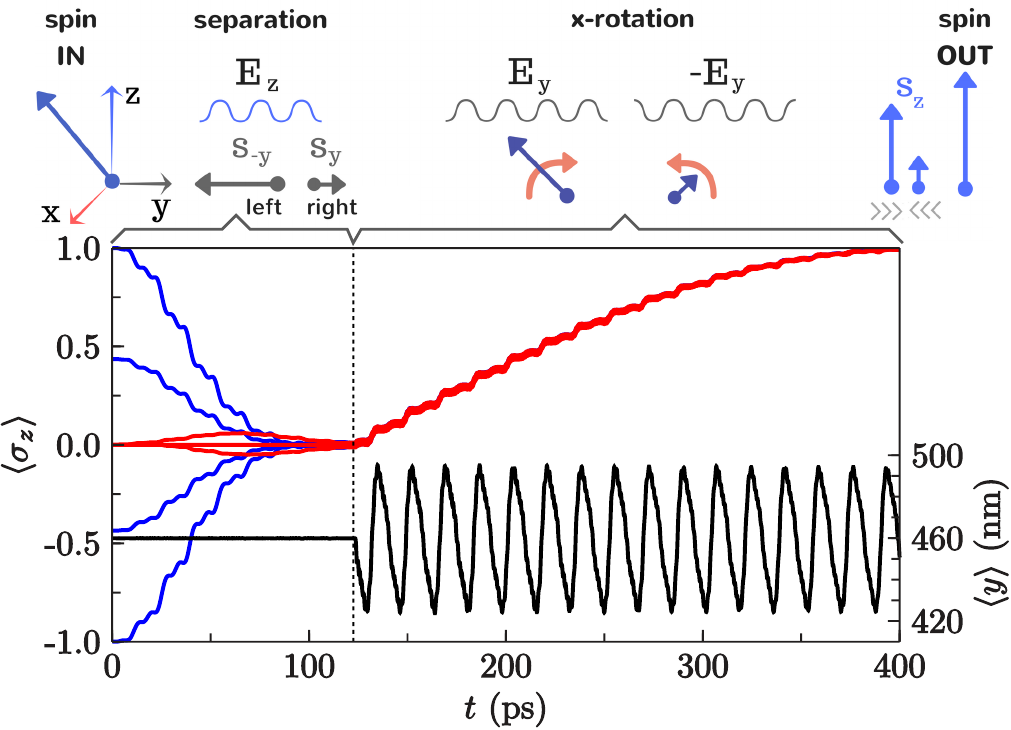}
\caption{\label{fig:initialization} Expectation value of the spin $z$-component $\langle\sigma_z\rangle$ during the initialization for different initial spin orientations lying in the $yz$ plane (blue curves) and in the $xy$ plane (red curves). Arbitrary spin is first spatially separated into two parts with spin parallel and antiparallel to the $y$-axis. Next, by opposite rotations of the left and right parts, spin is finally set along the $z$-axis.}
\end{figure}

We can now proceed to the next operation, namely, spin rotation about the $x$-axis. If we rotate the spin in the left half of the device clockwise about the $x$-axis, and in the right half counterclockwise, by $90^\circ$, both wavepacket parts will gain spins directed along the $z$-axis. The Hamiltonian (\ref{eq:soihamiltonian}) implies that motion in the $x$-direction results in spin rotation about the $y$-axis while motion in the $y$-direction rotates spin about the $x$-axis. Therefore we need to put the electron into motion along the $y$-axis. The voltages applied to the rails $U^\mathrm{rail}$ stabilizes the wavepacket in the middle between them. Fig. \ref{fig:layout} shows the charge density at the very beginning of the simulation. We can induce small oscillations of the electron in the $y$-direction by applying sinusoidal voltages $U_{\mathrm{as}}(t)=U_{\mathrm{as}}^0\sin(\omega t)$ between upper $U^\mathrm{a,c}$ and lower $U^\mathrm{b,d}$ gates, introducing a potential asymmetry in this direction. The black curve in Fig. \ref{fig:initialization} shows the expectation value of the electron position along the $y$-axis, calculated in the right half of the nanodevice as
\begin{equation}
\langle{}y\rangle_\mathrm{right}=\frac{\int_{L_x/2}^{L_x}dx\int_{0}^{L_y}dy\mathbf{\Psi^\dagger}y\mathbf{\Psi}}{\int_{L_x/2}^{L_x}dx\int_{0}^{L_y}dy\mathbf{\Psi^\dagger}\mathbf{\Psi}}
\end{equation}
This is not a resonant process and the frequency $\omega$ can be arbitrary, yet not too high as the oscillations cease to be adiabatic which significantly reduces the fidelity of spin initialization. In our simulations we assumed $\omega=60\,\mathrm{GHz}$ ($\hbar\omega=0.24\,\mathrm{meV}$). This is the highest value for which the achieved spin initialization accuracy exceeded 99\%. Electron oscillations along a straight line are not enough to rotate the spin, since after every period the spin reverts to the initial value. This situation changes when the SOI coupling depends on time \cite{pawlowski1}. We can accomplish this by applying an additional oscillatory voltage, with a phase shift of $\pi/2$ with respect to $U_{\mathrm{as}}$, namely $U_{\mathrm{off}}(t)=U_{\mathrm{off}}^0\cos(\omega t)$ , to all the gates including $U^\mathrm{top}$. This way the electron moves forwards with a different coupling than when it moves backwards and the spin rotations gradually accumulate. The voltage $U_{\mathrm{off}}$ is identical for all the gates. On the other hand $U_{\mathrm{as}}$ must be applied in a way, so that spin in the left and right halves of the nanodevice rotate in opposite directions. The values of $\langle{}y\rangle_\mathrm{right}$ and analogous $\langle{}y\rangle_\mathrm{left}$ oscillate with opposite phases as shown in the diagram above Fig. \ref{fig:initialization}. We thus apply the following voltages:
\begin{align}
U^\textrm{top,rail}(t)&=U_\textrm{bias}+U_{\mathrm{off}}(t),\\
U_0(t)&=U_\textrm{bias}+U_{\mathrm{off}}(t)+U_{\mathrm{barr}},\\
U_i^\mathrm{a,d}(t)&=U_\textrm{bias}+U_{\mathrm{off}}(t)+U_{\mathrm{as}}(t)+i^2U_\mathrm{par},\\
U_i^\mathrm{b,c}(t)&=U_\textrm{bias}+U_{\mathrm{off}}(t)-U_{\mathrm{as}}(t)+i^2U_\mathrm{par}
\end{align}
with $U_\textrm{bias}=-500\,\mathrm{mV}$ (new bias voltage), $U_{\mathrm{as}}^0=250\,\mathrm{mV}$, $U_{\mathrm{off}}^0=250\,\mathrm{mV}$, $U_{\mathrm{par}}=-5\,\mathrm{mV}$, $U_{\mathrm{bar}}=-875\,\mathrm{mV}$.
Please note the signs in front of $U_{\mathrm{as}}$.
Finally, Fig. \ref{fig:initialization} shows the spin $z$-component $\langle\sigma_z\rangle$ evolutions for several different initial spin orientations. The blue curves correspond to spins lying in the $yz$ plane, while the red ones in the $xy$ plane. The evolutions differ only in the first phase of the device operation, lasting about $125\,\mathrm{ps}$. After this time the wavepacket is separated into two parts of definite spins. One is parallel to the $y$-axis and the other antiparallel. At this moment the spin $z$-component equals zero. In the latter phase of operation, in which the spins are being rotated, the courses of $\langle\sigma_z\rangle$ overlap regardless of the initial spin orientation. At $t=403\,\mathrm{ps}$ they all reach values close to 1. At this moment we merge both wavepacket parts removing the potential barrier that separates them spatially. To do this we again put $U_\mathrm{barr} = 0$. The barrier should not be removed instantly and it is worth to spend a dozen of picoseconds for this part to avoid unnecessary energy rise of the final electron state, resulting in wavepacket oscillations. 

The final value of spin slightly depends on the initial spin orientation but for all the simulated cases the fidelity falls between $99.3\%$ and $99.7\%$. According to our calculations the highest fidelity is obtained for initial states being equally weighted linear combinations of spins parallel and antiparallel to the $y$-axis, hence for initial spin parallel to the $z$ or $x$ axis. The lowest fidelity has been obtained for highly non-equal linear combination weights. It is possible to decrease the frequency of oscillations below $50\,\mathrm{GHz}$ for the price of a proportional increase of the initialization time. This way, however, the fidelities may rise.

Thus far, we managed to perform a simulation of spin initialization in time sufficiently lower than the spin coherence time only for InSb material parameters, for which the SO coupling is very strong: $\alpha_\mathrm{SO} = 523\,\text{\AA}^2$. Our simulations have been performed for an ideal QW without any doping or structural imperfections. Currently fabricated InSb QWs are not yet ideal and their imperfection can result in randomly fluctuating fields giving contributions to the Rashba SO interaction \cite{bindel}. The electron wavefunction in our nanodevice is spatially stretched over a distance of about $d=600\,\mathrm{nm}$ along the $x$-axis, which is much greater than the fluctuation correlation length $\lambda=30\,\mathrm{nm}$. Therefore, their influence is effectively averaged (proportionally to $\lambda/d$) and significantly reduced. Because these fluctuations are time independent (do not depend on gate voltages), their
influence on the second stage of the simulation, namely on spin rotations, is negligible. However the first stage, in which two wavepacket parts change their positions, might be disturbed and the fidelity of spin initialization reduced. It should be possible to avoid such fluctuations in nanostructures made of Si/SiGe. In Si the Rashba coupling $\alpha_\mathrm{SO}$ is about 5000 times lower than in InSb \cite{tahan} but the coherence time is several orders longer \cite{takeda,pla,laucht} and the electron effective mass is many times higher, which gives hope that the initialization time will be sufficiently lower than the coherence time.

The nanodevice presented in Figs. \ref{fig:structure} and \ref{fig:layout} might be multiplied by putting copies of the device to the left and to the right. This way we obtain a multiple QD. However, if we only duplicate the nanodevice we can confine two electrons inside, then set the spins of each one of them separately and merge. This way also singlet-triplet qubits can be initialized.

We have designed a nanodevice capable of initializing spin of a single electron regardless of its initial orientation. By applying a sequence of voltages to the local gates, we can set the spin parallel to the $z$-axis. The process does not depend on the initial spin and the outcomes are virtually identical. The operation of the device does not require any external fields, microwaves or photons. The goal is achieved \emph{all-electrically} through application of voltages to the gates. The calculations are done for realistic material constants and the assumed geometry details. Our approach includes the fundamental electrostatic effects important for operation of the device. The performed simulations take into account subtle effects not accounted for using perturbation methods and model potentials. Interface imperfections have not been taken into account. Short range interactions are
averaged due to spatial stretching of the wavefunction along the $x$-axis over a distance of about $600\,\text{nm}$. However, charges trapped at the interfaces can be a source of potential nonparabolicity, which can affect the operation of the nanodevice.

\begin{acknowledgments}
This work has been supported by National Science Centre, Poland, under UMO-2014/13/B/ST3/04526. 
\end{acknowledgments}

\bibliography{thebibliography}

\end{document}